\documentclass[a4paper]{jpconf}
\usepackage{hyperref}
\usepackage{graphicx}
\begin{document}
\title{Density Estimation Trees as fast non-parametric modelling tools}

\author{Lucio Anderlini}

\address{Istituto Nazionale di Fisica Nucleare, Sezione di Firenze -- via Sansone 1, Sesto Fiorentino, 50019 Italy}

\ead{Lucio.Anderlini@cern.ch}

\begin{abstract}
  A Density Estimation Tree (DET) is a decision trees trained on a multivariate 
  dataset to estimate the underlying probability density function. While not competitive 
  with kernel techniques in terms of accuracy, DETs are incredibly fast, 
  embarrassingly parallel and relatively small when stored to disk. These 
  properties make DETs appealing in the resource-expensive horizon of the LHC 
  data analysis. Possible applications may include selection optimization, fast 
  simulation and fast detector calibration.
  In this contribution I describe the algorithm and its implementation made 
  available to the HEP community as a RooFit object.  A set of applications under 
  discussion within the LHCb Collaboration are also briefly illustrated.
\end{abstract}

\section{Introduction}
The fast increase of computing resources needed to analyse the data 
collected in modern hadron-collider experiments, and the lower cost 
of processing units with respect to storage, pushes High-Energy Physics (HEP)
experiments to explore new techniques and technologies to move as much
as possible of the data analysis at the time of the data acquisition 
(\textit{online}) in order to select candidates to be stored on disk,
with maximal, reasonably achievable, background rejection.
%
%Besides, the complexity of the data analyses and the important 
%inputs that the HEP community is receiving from the fast-growing 
%community of \emph{Data Scientists} motivate research and studies 
%of multivariate algorithms able to operate in the distributed computing
%environments, being today key elements in data processing and analysis.
%
Besides, research on multivariate algorithms, active both within and
outside of the HEP community, is approaching the challenge of operating 
in distributed computing environments, which represents a further motivation for studies of
new classes of algorithms.

Statistical inference of probability density functions underlying 
experimental datasets is common in \emph{High Energy Physics}.
\emph{Fitting} is an example of \emph{parametric}
density estimation. When possible, defining a parametric form of
the underlying distribution and choose the values for the parameters
maximizing the likelihood is usually the best approach.
In multivariate problems with a large number of variables and important
correlation, however, fitting may become unpractical, and \emph{non-parametric
density estimation} becomes a valid, largely employed, solution.

In HEP, the most common non-parametric density estimation, beyond the histogram, is probably
\emph {kernel density estimation} \cite{Cranmer:2000du}, based on the sum of 
normalized kernel functions centered on each data-entry.

In this write-up, I discuss \emph{Density Estimation Trees}, algorithms
based on a \emph{multivariate, binary tree} structure,
oriented to \emph{non-parametric density estimation}.
Density Estimation Trees are less accurate than kernel density estimation,
but much faster.
Integrating Density Estimation Trees is also trivial and fast, making 
iterative search algorithms convenient. 
Finally, storing a Density Estimation Trees and propagate it through 
the computing nodes of a distributed system is relatively cheap, offering a
reasonable solutions for compressing the statistical information
of large datasets.

An implementation of the algorithm in ROOT/RooFit is available through
CERN GitLab\footnote{\href{https://gitlab.cern.ch/landerli/density-estimation-trees}{gitlab.cern.ch/landerli/density-estimation-trees}}.

%This document is organized as follows, Section \ref{sec:algorithm} describes
%the algorithm and a RooFit \cite{} implementation made available through CERN 
%GitLab
%Section \ref{sec:applications} describes future applications in the framework of 
%the LHCb experiment.
%Finally, Section \ref{sec:summary} is devoted to a brief summary and outlook.

\section{The algorithm}\label{sec:algorithm}
  The idea of iteratively splitting a data sample, making the density 
  estimation to coincide with the average density in each portion of the data 
  space is not new.
  However, the technique had little room for applications in analyses 
  of datasets up to a few thousands of entries described by small sets 
  of correlated variables.
  
  Recently, $kd$-trees \cite{kdtrees} have been used to split large samples 
  into sub-sets consisting of equal fractions of the data entries. % \cite{}. 
  The idea underlying $kd$-trees is the iterative splitting of the data-sample
  using as threshold the median of a given projection. 
  While powerful to solve a vast range of problems, 
  including notably nearest-neighbour searches, 
  the lack of appropriateness of $kd$-trees to estimate probability densities
  is evident considering samples including multiple 
  data-entries.

  Ideally, the density estimation $\hat f(\mathbf{x})$ approximating the underlying 
  density function $f(\mathbf{x})$ should minimize the quantity
  $%\begin{equation}
    \mathcal R = \int \big( \hat f(\mathbf {x}) - f(\mathbf {x}) \big)^2 \mathrm{d}\mathbf{x}.
  $ %\end{equation}
  It can be shown \cite{dets} that, exploiting the Monte~Carlo approximation 
  \begin{equation}
    \lim_{N \to + \infty} \frac{1}{N} \sum_{i=1}^{N} g(x_i) = \int g(x)h(x) \mathrm d x \qquad \mbox{where } x_1, x_2, ..., x_N \mbox{ distribute as } h(x),
  \end{equation}
  the minimization of $\mathcal R$ is equivalent to growing a 
  Density Estimation Tree iteratively splitting the node $\ell$,
  into the two sub nodes $\ell_R$ and $\ell_L$ 
  minimizing the \emph{Gini} index,
   $ G(\ell) = R(\ell) - R(\ell_R) - R(\ell_L)$.
  Here, $R(\ell)$ represents the \emph{replacement error}, defined by 
  \begin{equation}
    R(\ell) \equiv - \frac{N_\ell^2}{N^2_\mathrm{tot} V_\ell},
  \end{equation}
  where $V_\ell$ is the hyper-volume of the portion of the data-space 
  associated to the node $\ell$, and $N_\ell$ the number of data
  entries it includes;
  $N_\mathrm{tot}$ is the number of data entries in the whole dataset.

\begin{figure}
    \begin{minipage}{0.5\textwidth}{}
      \includegraphics[width=\textwidth]{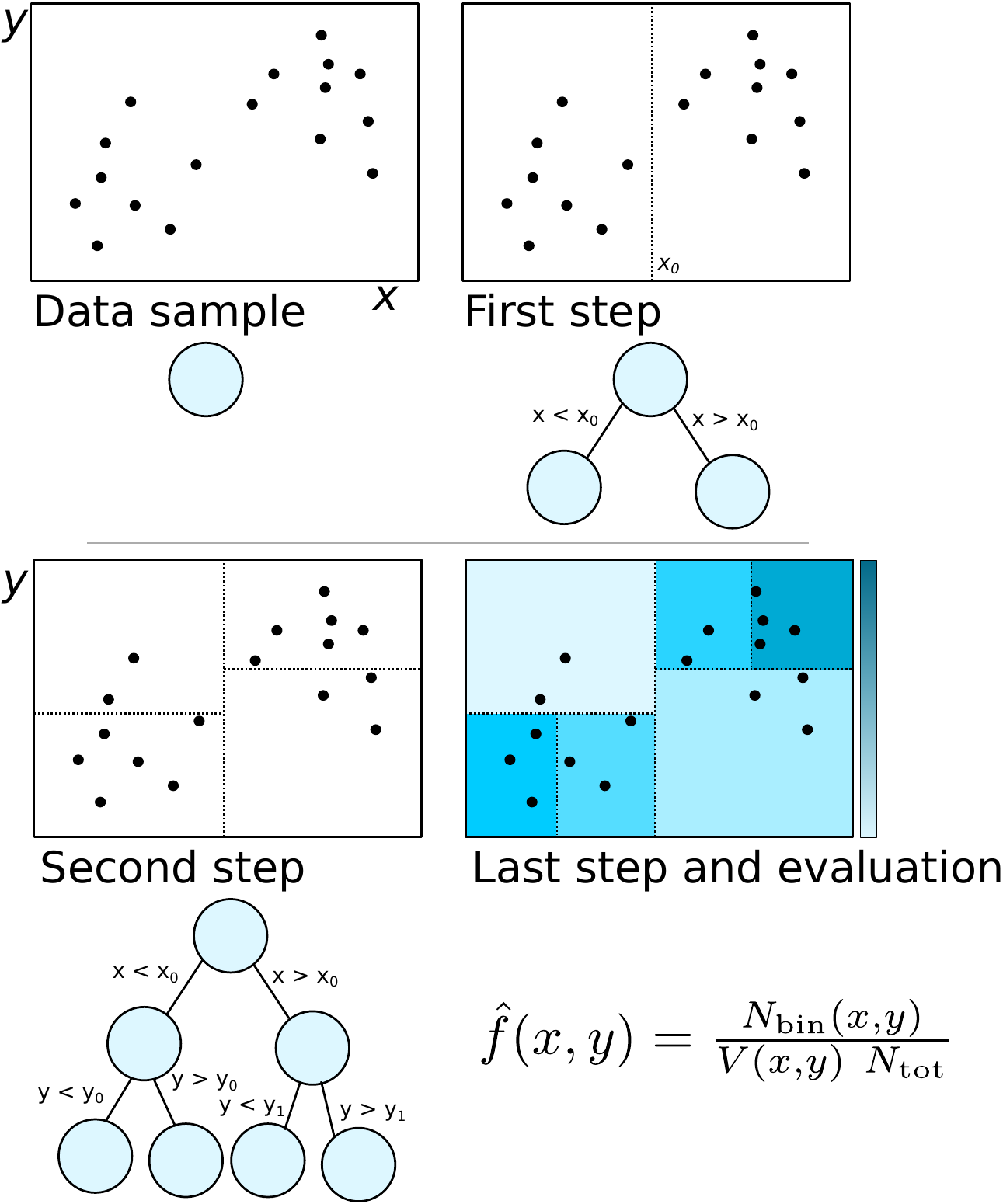}
      \caption{\label{fig:training} Schematic representation of the training 
        and the evaluation 
        procedures of a Density Estimation Tree.}
    \end{minipage}
    \hfill
    \begin{minipage}{0.45\textwidth}{}
      %\null\hfill\includegraphics[width=0.7\textwidth]{img/kdTree.png}\\[1mm]
    %  \null\hfill\includegraphics[width=0.7\textwidth]{img/detWiki.png}
    %  \caption{\label{fig:example} Density Estimation Tree trained on a 
    %  small two-dimensional dataset containing entries (), (), ()...
    %  . The colors represent the density 
    %  estimation blue tones correspond to the lowest densities, 
    %  red one to the highest.}
      \includegraphics[width=.9\textwidth]{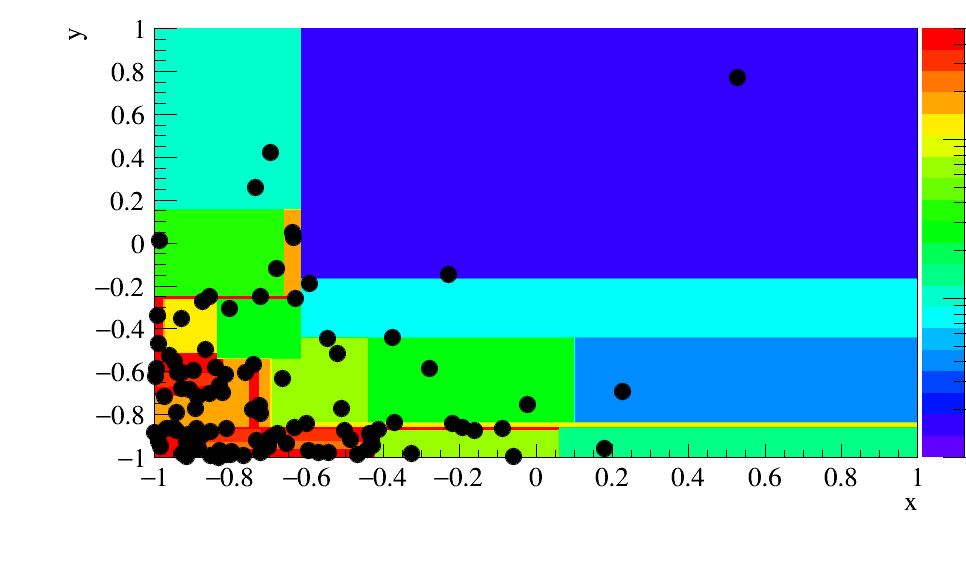}\\
      \includegraphics[width=.9\textwidth]{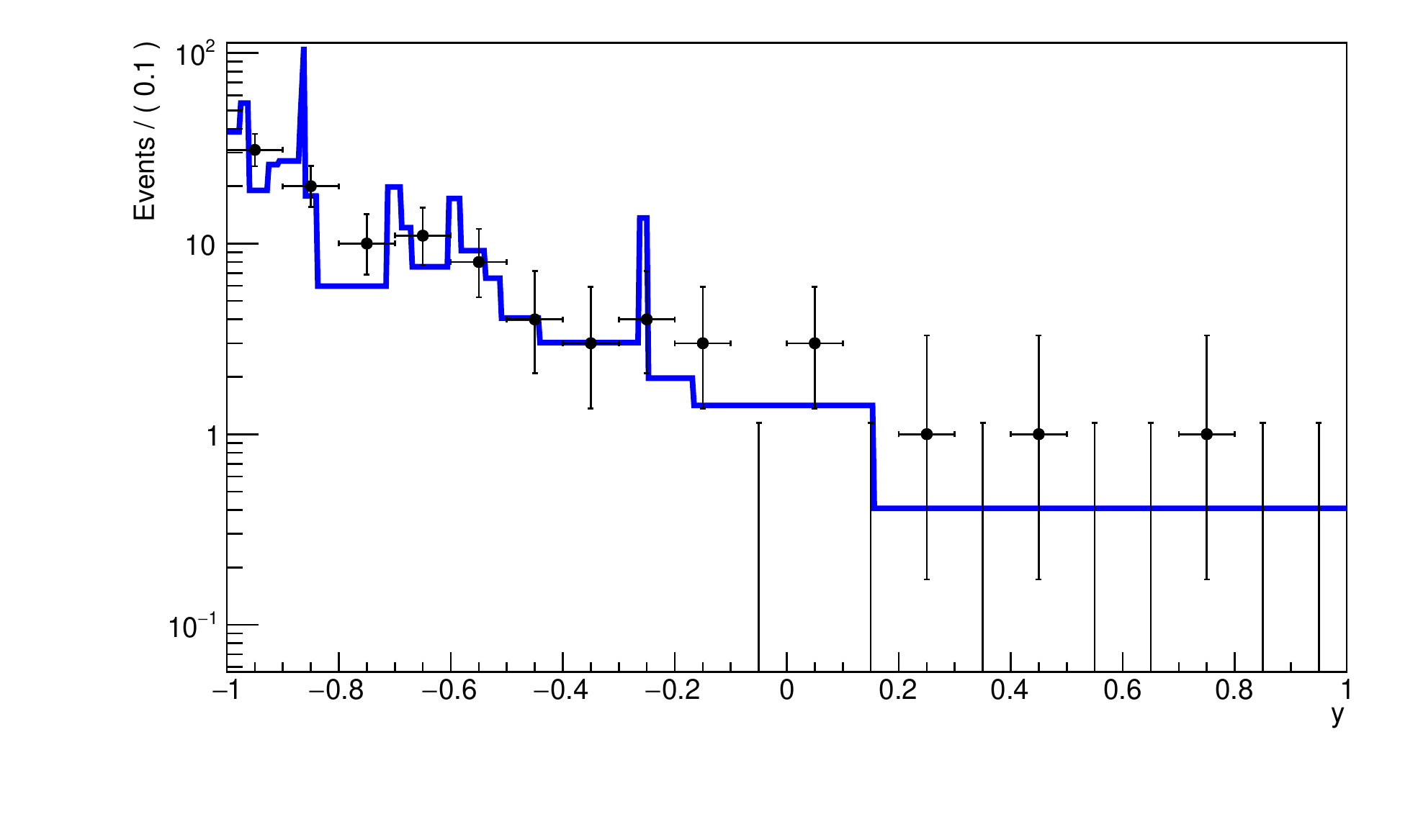}
      \caption{\label{fig:overtraining} Top, an example of an overtrained 
        Density Estimation Tree. The random alignment of data-entries 
        with respect to one of the variables describing the problem 
        is misinterpreted as a spike in the density estimation, as 
        evident in the projection onto the vertical axis shown on the bottom.
      }
    \end{minipage}
  \end{figure}

  Figure \ref{fig:training} shows an example of how the training is performed.

  \subsection{Overtraining}
    As in the case of Classification algorithms, overtraining is the 
    misinterpretation of statistical fluctuations of the dataset as 
    relevant features to be reproduced by the model.

    An example of overtraining of Density Estimation Trees is presented
    in Figure \ref{fig:overtraining}. In the presented dataset, the 
    alignment of data-entries in one of the input variables is interpreted
    as narrow spikes. To compensate spikes, in terms of absolute normalization,
    the density is underestimated in all of the other points of the 
    parameter space.

    %Overtraining can be fought using either techniques developed specifically
    %for decision trees or for density estimation algorithms.
    Overtraining in decision trees is controlled through an iterative approach 
    consisting in \emph{pruning} and \emph{cross-validation}: finding and 
    removing the branches increasing the complexity of the 
    tree without enhancing the statistical agreement with a set of test
    samples.
    Cross-validation is very expensive in terms of computing power    
    and often fails to identify problems of over-training arising close to 
    the root of the decision tree.
    
    Overtraining in density estimation is fought, instead, by defining 
    \emph{a priori} an expected resolution width, neglecting fluctuations
    under that resolution while building the statistical model.
    For example, kernel density estimation algorithms require 
    a parameter, named \emph{bandwidth} as an input.
    The bandwidth is related to the width
    of the kernel function. Abundant literature exists on techniques 
    to optimise the bandwidth for a certain dataset,
    most of them represent a preliminary step of the density estimation algorithm.
    Growing a Density Estimation Tree with a minimal leaf width is 
    fast, doesn't require post-processing and it is found to result in  
    better-quality estimations with respect to cross-validation 
    procedures.
    The same techniques used to compute the optimal 
    bandwidth parameter for kernel density estimation algorithms can 
    be used to define the optimal minimal leaf width
    of the Density Estimation Tree.

    %\begin{figure}
    %\end{figure}

  %\subsection{Smoothing}
  %  Kernel Density Estimation is sometime adopted to smooth histograms
  %  by sampling them and then building the kernels. The procedure is 
  %  known as \emph{kernel smearing} \cite{} and reflects the need, for specific
  %  applications, of smooth estimations of the probability density function.
  %  
  %  A cheaper alternative to kernel smoothing can be obtained 
  %  convolving the estimation obtained with trees with a resolution function.
  %  Choosing a triangular (or \emph{tent}) resolution function 
  %  reduces the computation of the convolution to the algebraical sum of 
  %  few terms, making it fast and reliable. 

  \subsection{Integration}
    As mentioned in the introduction, fast integration of the statistical 
    model built using Density Estimation Trees is one of the strengths of 
    the algorithm.
    Integration usually responds to two different needs: \emph{normalization}
    and \emph{slicing} (or \emph{projecting}, or \emph{marginalizing}).

    Integrals to compute the overall normalization of the density estimation,
    or the contribution in a large fraction of the data-space,
    gain little from exploiting the tree
    structure of the density estimator. A sum over the contributions of each
    leaf represents the best strategy.

    Instead, integrals over a narrow subset of the data-space should 
    profit of the tree structure of the density estimation to exclude
    from the integration domain as many leaves as possible, as early
    as possible. Exploiting the tree structure when performing integrals 
    of slices 
    can drastically reduce the computing time in 
    large density estimation trees.

  \subsection{Operations with Density Estimation Trees}
    Combining weighted Density Estimation Trees can be useful 
    to model data samples composed of two or more components.
%    Density Estimation Trees can be used, for example,
%    to model data samples including different components. 
%    It is therefore useful to combine different Decision 
%    Trees, with different weights, into a unique Density 
%    Estimation Tree.
% 
%    To achieve combination of Decision Trees, one needs 
%    at least the capability of performing scalar operations 
%    (such as multiplying by a constant) 
%    and binary operations (such as summing two Density Estimation Trees).
%    The implementation of the former is pretty trivial, since it is sufficient
%    to apply the scalar operation to value contained in each node
%    composing the Density Estimation Tree, therefore here I focus on 
%    the latter.
%
    Combination is achieved implementing both scalar and binary 
    operations. There is not much to discuss about scalar operations, 
    where the scalar operation is applied to each leaf independently. 
    Instead, binary operations require the combination of two different Density Estimation 
    Trees, which is not trivial because the 
    boundaries are \emph{a priori} different. The algorithm to combine two
    Density Estimation Trees consists of the iterative splitting of the terminal
    nodes of the first tree, following the boundaries of the terminal
    nodes of the second one. 
    Once the combination is done, the first tree is compatible with the second one
    and the binary operation can be performed 
    node per node.
    The resulting tree may have several additional
    layers with respect to the originating trees, therefore a final step
    removing division between negligibly different nodes 
     is advisable.

\section{Discussion of possible applications}\label{sec:applications}
  Density Estimation Trees are useful to approach problems defined by
  many variables and for which huge statistical samples are available.
  To give a context to the following examples of applications, I consider
  the calibration samples for the Particle IDentification (PID) algorithms at
  the LHCb experiment.

  PID calibration samples are sets of decay candidates reconstructed and 
  selected relying on kinematic variables only, to distinguish between
  different types of long-lived particles: electrons, muons, pions, kaons,
  and protons.
  
  The PID strategy of the LHCb detector relies on the combined response 
  of several detectors: two ring Cherenkov detectors, an electromagnetic 
  calorimeter, a hadronic calorimeter and a muon system \cite{lhcb}.
  The response of the single detectors are combined into likelihoods used 
  at analysis level to define the tightness of the PID requirements. 

  Calibration samples count millions of background-subtracted candidates,
  each candidate is defined by a set of kinematic variables, for example 
  momentum and pseudorapidity, and a set of PID likelihoods, one per 
  particle type.
  The correlation between all variables is important and not always
  linear.

  \subsection{Efficiency tables}
    The first application considered is the construction of tables defining 
    the probability that the PID likelihood of 
    a candidate, defined by a set of kinematic 
    variables, satisfies a particular requirement.
    
    Building two Density Estimation Trees with the kinematic variables
    defining the data-space, one with the full data sample 
    (tree $t_\mathrm{all}$), and one 
    with the portion of data sample passing the PID criteria 
    (tree $t_\mathrm{pass}$), allows to compute the efficiency for 
    each combination of the kinematic variables by evaluating the 
    Density Estimation Tree obtained taking the ratio $t_\mathrm{pass}/t_\mathrm{all}$.

    For frequently-changing criteria a dynamic determination of the efficiency can 
    be envisaged. For simplicity, consider the generic univariate PID criterion $y > 0$.
    In this case a single Density Estimation Tree $d (x_1, x_2, y)$ 
    defined by the kinematic variables
    $(x_1, x_2)$, and one PID variable $y$, has to be trained on the calibration sample.
    The dynamic representation of the efficiency for a candidate 
    having kinematic variables $(\hat x_1, \hat x_2)$ is the ratio
    \begin{equation}
      \epsilon (x_1, x_2; y > 0) =  {\displaystyle \int_{y>0} d ( \hat x_1, \hat x_2 , y ) \mathrm dy }\Bigg/{\displaystyle \int_{\mathrm{any\ }y} d ( \hat x_1, \hat x_2 , y ) \mathrm dy}.
    \end{equation}

    Thanks to the fast slice-integration algorithm, the computation of this ratio
    can be included in an iterative optimization procedure aiming at 
    an optimization of the threshold on $y$.

  \subsection{Sampling as fast simulation technique}
    Another important application is related to fast simulation of HEP events.
    Full simulation, including interaction of the particles with matter,
    is becoming so expensive to be expected 
    exceeding the experiments' budgets in the next few years.
    Parametric simulation is seen as a viable solution, as proved by the 
    great interest raised by the DELPHES project \cite{deFavereau:2013fsa}.
    However, parametrizing a simulation presents the same pitfalls as parametrizing  
    a density estimation: when correlation among different variables becomes
    relevant, the mathematical form of the parametrization increases in 
    complexity up to the point it becomes unmanageable. 

    Density Estimation Trees are an interesting candidate for non-parametric
    fast simulation. Let $d(x_1, ..., x_N, y_1, ..., y_n)$ be a Density 
    Estimation Tree trained on a set 
    of candidates defined by \emph{generator} variables $\mathbf x \equiv (x_1, ..., x_N)$, and 
    by variables $\mathbf y \equiv (y_1, ..., y_n)$ obtained through full simulation.
    For example, $\mathbf x$ could represent the kinematic variables of a track and 
    $\mathbf y$ the PID likelihoods. 

    The aim of fast non-parametric simulation is, given a new set of values
    $(\hat x_1, ..., \hat x_N)$ for $\mathbf x$, to compute a set of values for 
    $\mathbf y$ distributed according to the conditional probability density function
    $d(y_1, ..., y_n|\hat x_1, ..., \hat x_N)$.

    Once the DET is trained, 
    the tree structure of the density estimator is used to compute for each leaf $\ell$ 
    the hyper-volume $V_{\ell \cap \hat \mathbf{x}}$ of the intersection between $\ell$ 
    and hyper-plane defined by
    $\mathbf x = (\hat x_1, ..., \hat x_N)$.
    
    A random leaf $L \in \{\ell\}$ is then chosen with probability proportional 
    to $V_{L \cap \hat \mathbf{x}}$, and variables $\mathbf y$ are generated
    following a flat distribution bounded within $L$.

    A set of \emph{generator} variables $\mathbf{x}$ can then be completed by
    the corresponding $\mathbf{y}$ variables without full simulation, but 
    relying on the joint multivariate distribution learnt 
    by the Density Estimation Tree.

\section{Summary and outlook}\label{sec:summary}
  I discussed Density Estimation Tree algorithms as fast modelling tools for 
  high statistics problems characterized by a large number of correlated variables
  and for which an approximated model is acceptable.
  The fast training and integration capabilities make these algorithms of interest 
  for the high-demanding future of the High-Energy Physics experiments. 
  The examples discussed, which benefited from an active discussion within the 
  Particle Identification Group of the LHCb collaboration, explore cases where 
  the statistical features of huge samples have to be assessed in a time shorter than what
  standard estimators would require.
  In future, Density Estimation Trees could be sampled to train Regression Multivariate
  Algorithms, such as Neural Networks, in order to smooth the response and 
  further speed up the query time,
  at the cost of loosing its fast-integration properties.

\section*{Acknowledgements}
  I thank Alberto Cassese, Anton Poluektov, and Marco Cattaneo 
  for the encouragements in developing this work and for the useful discussions
  we had.
    
\section*{References}
%\cite{Cranmer:2000du}

\end{document}